\documentclass[aps,pra,groupedaddress,showpacs]{revtex4-1}
\usepackage{graphicx}
\usepackage{amsmath}
\usepackage{textcomp}
\usepackage{mathrsfs}
\usepackage{amsfonts}
\usepackage{mathtools}
\DeclarePairedDelimiter{\ceil}{\lceil}{\rceil}

\begin{document}

\newcommand{\beq}{\begin{equation}}
\newcommand{\eeq}{\end{equation}}
\newcommand{\barr}{\begin{eqnarray}}
\newcommand{\earr}{\end{eqnarray}}
\newcommand{\bseq}{\begin{subequations}}
\newcommand{\eseq}{\end{subequations}}
\newcommand{\bal}{\begin{align}}
\newcommand{\eal}{\end{align}}
\newcommand{\ket}[1]{|#1\rangle}
\newcommand{\bra}[1]{\langle #1|}
\newcommand{\expectation}[3]{\langle #1|#2|#3 \rangle}
\newcommand{\braket}[2]{\langle #1|#2\rangle}
\newcommand{\vett}[1]{\mathbf{#1}}
\newcommand{\uvett}[1]{\hat{\vett{#1}}}
\newcommand{\uvettGreek}[1]{\hat{\boldsymbol{#1}}}
\newcommand{\vettGreek}[1]{\boldsymbol{#1}}

\title{Vector Properties of Radially Self-Accelerating Beams}

\author{Marco Ornigotti$^{1,*}$}
\author{Alexander Szameit$^1$}
\affiliation{$^1$Institut f\"ur Physik, Universit\"at Rostock, Albert-Einstein-Stra\ss e 23, 18059 Rostock, Germany}

\email{marco.ornigotti@uni-rostock.de}

\date{\today}

\begin{abstract}
We present a complete and consistent theory of vector radially self-accelerating beams (RSABs). We use this theory as a model for describing the properties of focussed RSABs, and to show, in particular, that only circular polarised RSABs maintain their self-accelerating character upon focussing. Moreover, we also calculate the linear and angular momentum for paraxial vector RSABs, and discuss both their global and local properties.
\end{abstract}

\maketitle

\section{Introduction}
Accelerating beams, i.e., electromagnetic fields that propagate along curved trajectories in free space without being subject to any external force, have been the subject of a thorough investigation in the last years. The most famous representative of such class of beams is, without doubts, the Airy beam. Firstly introduced in the context of quantum mechanics by Berry and Balazs as an exotic solution of the Schr\"odinger equation \cite{berryAiry}, it was then introduced in optics in 2007 by Siviloglou and co-workers \cite{siviloglou,siviloglou2}, as an exact solution of the paraxial equation propagating along a parabolic trajectory in free space. Due to their intriguing features, Airy beams were studied in different contexts, such as nonlinear optics \cite{airy1}, particle manipulation \cite{airy2}, and proposed as an efficient way to generate curved plasma channels \cite{airy3}. 

Inspired by these results, the last years witnessed the emergence of many different types of accelerating beams in different coordinate systems, such as parabolic \cite{acc1} and Weber \cite{acc2} beams. Moreover, beams of light capable to propagate along curved \cite{acc3,curved1, curved2} and arbitrary \cite{arbitrary1,arbitrary2} trajectories, has also been proposed. Recently, two new classes of accelerating beams have been introduced, namely angular \cite{angularAcc, vettiOE} and radially self-accelerating beams \cite{nostroPRL, nostroAPL}. While the former acquire angular acceleration during rotation around their optical axis \cite{angularAcc},  the latter exhibit radial acceleration, a feature which makes them propagate along spiralling trajectories around their optical axis. 

Radially self-accelerating beams (RSABs) can be understood in terms of superpositions of Bessel beams, where each single component is characterised by an angular velocity proportional to the amount of orbital angular momentum it carries. This, ultimately, results in an electromagnetic field, whose transverse field or intensity distribution rotates around the propagation direction with a given constant angular velocity $\Omega$ \cite{nostroPRL}. Among the vast zoology of RSABs, in particular, helicon beams, i.e., a subclass of RSABs consisting of rotating diffraction-free beams based on the superposition of two Bessel beams with opposite orbital angular momentum, have attracted a lot of interest in the last decades \cite{nostroAPL,helicon1,helicon2,helicon3,helicon4,helicon5,helicon6,helicon7,helicon8,helicon9}. Beyond helicon beams, RSABs have potentially significant  applications in different areas of physics, such as sensing \cite{airy3}, material processing \cite{matProc1,matProc2}, and particle manipulation \cite{partMan1,partMan2}. 

Despite this broad interest, RSABs have only been defined within the scalar electromagnetic theory, and their vector nature, as well as the effect of focussing on their self-accelerating character, has not been yet investigated. In this work, therefore, we introduce vector RSABs, and study their vector properties, in terms of their linear and angular momentum content. Moreover, we carefully analyse what is the impact of focussing on the self-accelerating character of RSABs, and under which conditions the focussing process does not spoil this property.

This work is organised as follows: in Sect. 2 we briefly recall the definition of RSABs, and recall some of their main properties. In Sect. 3 we use the method of Hertz potentials to construct vector RSABs, and use these solutions as a model for focussed RSABs. Then, we derive a condition on the polarisation that a scalar RSAB must possess, in order to maintain its self-accelerating character upon focussing. Section 4 is then devoted to calculate the linear and angular momentum for paraxial, intensity rotating RSABs.  Conclusions are then drawn in Sect. 5.

\section{Radially Self-Accelerating  Beams}
We start our analysis by considering scalar, monochromatic, free space solutions of the Helmholtz equation 
\beq\label{eq1}
\left(\nabla^2+k_0^2\right)\psi(\vett{r})=0,
\eeq
where $k_0$ is the vacuum wave vector. The most general solution of the above equation in cylindrical coordinates, can be given in terms of superposition of Bessel beams, i.e.,
\beq\label{eq2}
\psi(\vett{r})=\sum_{m}\,\int\,d\xi\,C_m(\xi)\text{J}_m(\rho\sqrt{1-\xi^2} )e^{i(m\theta+\xi\zeta)},
\eeq
where $\rho=k_0R$, and $\zeta=k_0 z$ are normalised radial and longitudinal coordinates, $\text{J}_m(x)$ is the Bessel function of the first kind \cite{nist}, and the integration variable $\xi=\cos\vartheta_0$  plays the role of the Bessel cone angle $\vartheta_0$ \cite{durnin}. 

From the above solution, it is possible to extract RSABs by applying the requirements that Eq. \eqref{eq2} must fulfil, in order to be a RSAB \cite{nostroPRL}. First, $\psi(\vett{r})$ must propagate freely, and not under the action of a certain potential. Then, there should exist a suitable reference frame, in which $\psi(\vett{r})$ is manifestly propagation invariant, i.e., no explicit $\zeta$-dependence must appear. Finally, an observer at rest in such reference frame should experience a fictitious force, which, ultimately, is at the core of self-accelerating character of RSABs.

Whle the first requirement is automatically met by the fact that we are considering free space propagation, the second one is very useful to define RSABs properly. Once it is fulfilled, in fact, it is not hard to show that the third requirement follows accordingly. We therefore require, that, after a suitable coordinate transformation $\vett{r}'=S\, \vett{r}$,  the field $\psi(\vett{r}')$ in the new coordinate frame is manifestly propagation invariant, i.e., 
$\partial\psi(\vett{r}')/\partial\zeta=0$. To this aim, we introduce the co-rotating coordinate $\Phi=\theta+\Lambda\zeta$, and choose the expansion coefficient as  $C_m(\xi)=D_m\delta(\xi-(m\Lambda+\beta))$, where $\Lambda=\Omega/k_0>0$ is the normalised angular velocity of the RSAB, and $\beta$ is a free (dimensionless) parameter, with the physical meaning of a normalised propagation constant. Substituting this Ansatz in Eq. \eqref{eq2} we get the following result
\beq\label{eq3}
\psi_{RSAB}(\rho,\Phi)=e^{i\beta\zeta}\sum_{m\in\mathcal{M}}D_m\text{J}_m(\alpha_m\rho)e^{im\Phi},
\eeq
where $\alpha_m=\sqrt{1-(m\Lambda+\beta)^2}$, and $\mathcal{M}=\{m\in\mathbb{N}: \alpha_m>0\}$. For $\beta=0$, the above field is manifestly propagation invariant, as no explicit $\zeta$-dependence is present. Moreover, its amplitude and phase both rotate with normalised angular velocity $\Lambda$ during propagation. For $\beta\neq 0$, on the other hand, the field itself is not anymore propagation invariant, due to the presence of the global phase factor $\exp{(i\beta\zeta)}$. Nevertheless, the intensity $|\psi_{RSAB}(\vettGreek{\rho})|^2$ is propagation invariant also for $\beta\neq 0$. In this case, however, while both intensity and phase propagate describing spiralling trajectories, they are not synchronised anymore. These two classes of RSABs are called field rotating, and intensity rotating, respectively \cite{nostroPRL}. 

\begin{figure}[!t]
\begin{center}
\includegraphics[width=\textwidth]{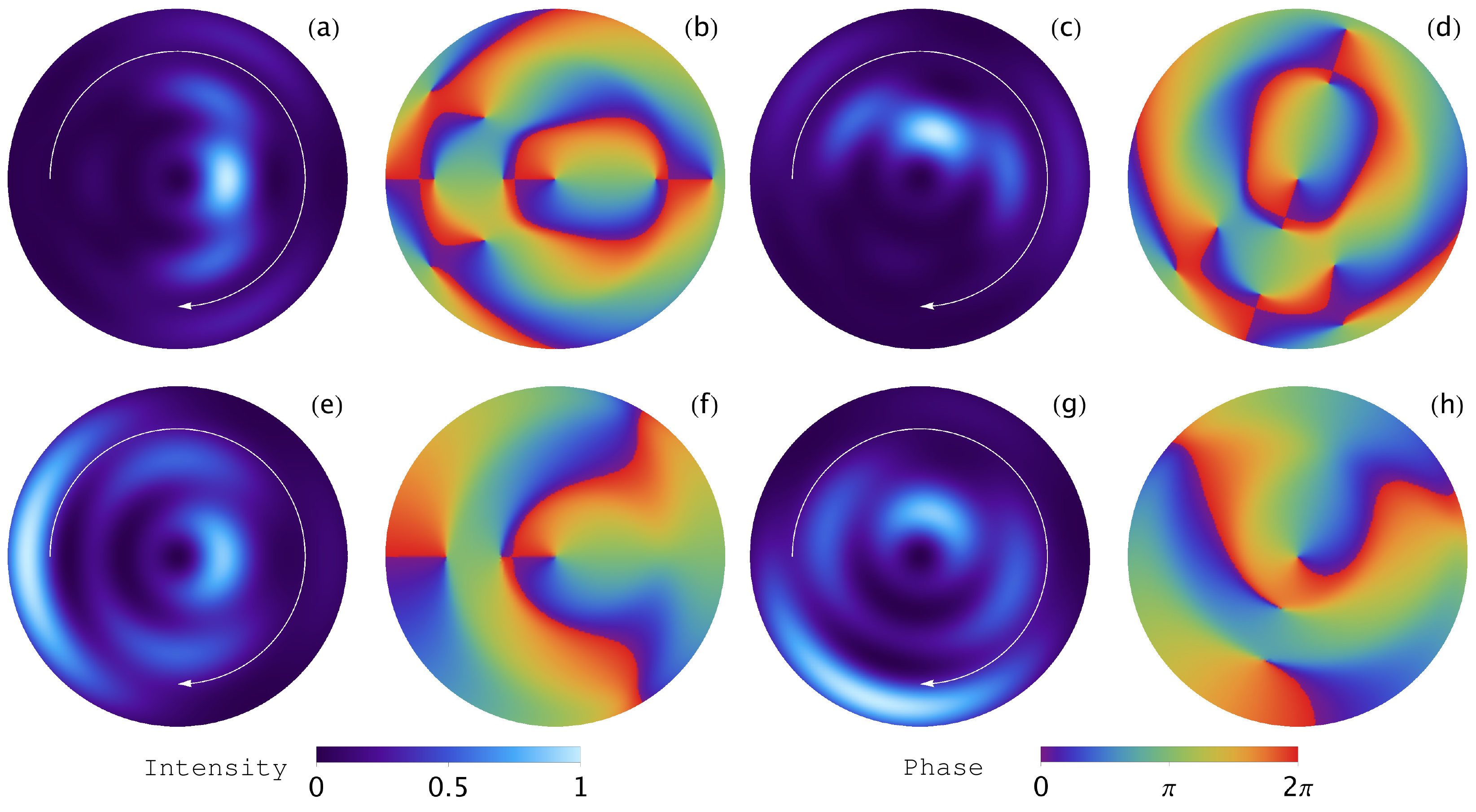}
\caption{Intensity and phase distribution for field rotating RSABs (top row) and intensity rotating RSABs (bottom row). Panels (a), (b), (e), and (f) correspond to the intensity and phase distributions at $z=0$, while panels (c), (d), (g), and (h) to $z=0.8(2\pi/\Lambda)$. Moreover, for the top row, the intensity and phase distributions have been plotted in the region $0\leq\rho\leq 10$, while for the lower row in the region $0\leq\rho\leq 1200$ has been chosen. The difference in the plotting range for the normalised radial coordinate $\rho$ reflects the paraxial (bottom) and nonparaxial (top) character of the plotted RSABs. In all these figures, $\Lambda=10^{-5}$ (corresponding to an angular velocity of $\Omega\simeq 75 $ rad/m at $\lambda=800$ nm), $m_{max}=4$, and $D_m=1$ has been used. For the top row, $\beta$ is set to zero, while for the bottom row $\beta=1-m_{max}\Lambda=0.99996$ (corresponding to a value of a global propagation constant $\beta_0\simeq 7.8$ $\mu m^{-1}$ for $\lambda=800$ nm) has been used. The white arrow in the intensity profiles show the direction of rotation.}
\label{figure1}
\end{center}
\end{figure}

It is worth noticing, moreover, that while for $\beta=0$ the set $\mathcal{M}$ contains only positive integers, for $\beta\neq 0$ positive and negative values of $m$ are allowed. Thus, helicon beams, for example, are a particular case of intensity rotating RSABs, where only two Bessel beams are participating in the sum in Eq. \eqref{eq3} An Example of both classes of RSABs is given in Fig. \ref{figure1}.

Of particular interest are RSABs with $\Lambda\ll 1$. Since $\Lambda=\Omega/k_0$, this condition corresponds to RSABs, whose actual angular velocity $\Omega$ is much smaller than the beam's wave vector $k_0$. This, ultimately, corresponds to experimentally realisable RSABs. In the rest of this manuscript, if not specified otherwise, we will always implicilty assume that $\Lambda\ll 1$ holds. This assumption, moreover, has different consequences for field and intensity rotating RSABs.
 
 In the former case (i.e., for $\beta=0$),  $\Lambda\ll1$ implies that the (normalised) transverse momentum of each Bessel component is given by $\alpha_m=\sqrt{1-m^2\Lambda^2}\simeq 1+\mathcal{O}(m^2\Lambda^2)$. If we recall, that the transverse momentum of a Bessel beam is related to the Bessel cone angle by the relation $k_{\perp}=k_0\sin\vartheta_0$, a value of the normalised transverse momentum $\alpha_m\simeq 1$ corresponds to $\vartheta_0\simeq\pi/2$, i.e., to a highly nonparaxial Bessel beam. 
 
 Despite this fact, however, the nature of the resulting RSAB can be tuned at will between paraxial and nonparaxial, by simply changing the number of Bessel beams that participate to the sum in Eq. \eqref{eq3}. To obtain nonparaxial RSABs, it is sufficient to limit the summation in Eq. \eqref{eq3} to $m_{max}=\text{max}\{\mathcal{M}\}<\ceil[\big]{\Lambda^{-1}}$. In this case, in fact, the transverse momentum of every Bessel component will be $\alpha_m\simeq 1$, and the resulting RSAB will be highly nonparaxial. 
 
 On the other hand, if one includes only values of $m$, that are close to $\ceil[\big]{\Lambda^{-1}}$, i.e., if $m\in[m_{max}-\bar{m},m_{max}]$ in Eq. \eqref{eq3}\footnote{with $\bar{m}$ small compared to $m_{max}$, such that $\bar{m}^2\Lambda^2\ll1$ still holds}, then $m\Lambda\simeq 1$, and, correspondingly, $\alpha_m\simeq 0$. In this case, all Bessel components will be paraxial (i.e., the correspondent cone angle will be $\vartheta_0\ll1$), and the resulting RSAB can also be interpreted as a paraxial beam.
 
 For intensity rotating RSABs (i.e., for $\beta\neq 0$), instead, $\alpha_m$ can be made arbitrarily small, independently from the value of $\Lambda$, by suitably tuning the parameter $\beta$. In this case, then, the paraxial limit is simply obtained by choosing $\beta$ such that $\alpha_m\simeq 0$, i.e., $\beta=1-m_{max}\Lambda$, with 
 $m_{max}=\text{max}\{\mathcal{M}\}$. Notice, that with this choice of $\beta$, $\alpha_{m_{max}}=0$, and therefore the sum in Eq.\eqref{eq3} extends to $m_{max}-1$, as $\text{J}_{m_{max}}(\alpha_{m_{max}}\rho)=0$. 
 
This extra flexibility in tuning the propagation constant $\beta$ and the angular velocity $\Lambda$ independently makes intensity rotating RSABs easier to generate and manipulate experimentally, than their field rotating counterparts \cite{nostroPRL, nostroAPL}.

\section{Vector Radially Self-Accelerating Beams}\label{vectorialisation}
The solution presented in Eq. \eqref{eq3} describes scalar RSABs. In many situations, however, a scalar representation of the electromagnetic field is not enough to fully describe its properties. A typical example is the focussing of a beam of light by means of a thick lens. On the focal plane of the lens, in fact, the scalar approximation given by Eq. \eqref{eq3} would fail to describe the properties of a focussed RSAB, and a full vector theory should be instead employed. A simple way to retrieve a full vector solution of Maxwell's equations from a solution to the scalar Helmholtz equation is given by the method of Hertz potentials  \cite{stratton, joptHertz}. First, one defines the Hertz potential $\boldsymbol\Pi(\vett{r},t)=\psi(\vett{r})\exp{(-i\omega t)}\uvett{f}$, where $\uvett{f}$ is a suitable polarisation unit vector, and $\psi(\vett{r})$ is a solution of Eq. \eqref{eq1}. Then, the vector electric and magnetic fields can then be retrieved from $\vettGreek{\Pi}(\vett{r},t)$ as follows:
\bseq\label{eq5}
\begin{align}
\vett{E}(\vett{r},t) & =-\frac{\partial\boldsymbol\Pi(\vett{r},t)}{\partial t},\\
\vett{B}(\vett{r},t) & =\nabla\times\nabla\times\boldsymbol\Pi(\vett{r},t).
\end{align}
\eseq
In the general case, both an electric ($\boldsymbol\Pi_e$) and magnetic ($\boldsymbol\Pi_m$) Hertz potential should be introduced, each accounting for the sources of electric and magnetic field, respectively. For free space propagation, however, no sources are present, and the electric and magnetic Hertz potential coincide (up to a global constant), i.e., $\boldsymbol\Pi_e=\boldsymbol\Pi_m\equiv\boldsymbol\Pi$ \cite{stratton}. According to the convention adopted by Jackson \cite{jackson}, the electric and magnetic fields defined by Eqs. \eqref{eq5} correspond to TE fields. The TM fields, however, can be obtained straightforwardly from the TE ones by setting $\vett{E}_{TM}\rightarrow\vett{B}_{TE}$, and $\vett{B}_{TM}\rightarrow-\vett{E}_{TE}/c^2$.

To calculate the vector electric and magnetic fields corresponding to RSABs, we first rewrite Eq. \eqref{eq3} as $\psi_{RSAB}(\vett{r})=\sum_mD_m\phi_m(\vett{r})$, with $\phi_m(\vett{r})=\exp{[i m\theta+i(m\Lambda+\beta)\zeta]}\text{J}_m(\alpha_m\rho)$ being the usual Bessel beam \cite{durnin}. This allows us to define the Hertz potentials for RSABs in terms of the Hertz potentials for ordinary Bessel beams, i.e.,
\beq\label{eq6}
\boldsymbol\Pi(\vett{r},t) =\sum_{m\in\mathcal{M}}D_m\vett{P}^{(m)}(\vett{r},t)
\eeq
where 
\beq\label{eq6bis}
\vett{P}^{(m)}(\vett{r},t)=\phi_m(\vett{r})e^{-i\omega t}\uvett{f}
\eeq
is the Hertz potential corresponding to a single Bessel beam $\phi_m(\vett{r})$, whose polarisation is defined by the unit vector $\uvett{f}$. Then, using Eq. \eqref{eq5}, we can first calculate the electric and magnetic vector fields for a single Bessel beam, namely
\bseq\label{eq8}
\begin{align}
\vett{E}^{(m)}(\vett{r},t) &=-\frac{\partial\vett{P}^{(m)}(\vett{r},t)}{\partial t},\\
\vett{B}^{(m)}(\vett{r},t) &=\nabla\times\nabla\times\vett{P}^{(m)}(\vett{r},t).
\end{align}
\eseq
Then, the electric and magnetic fields of vector RSABs can be written as follows:
\bseq\label{eq9}
\begin{align}
\vett{E}(\vett{r},t) & =\sum_{m\in\mathcal{M}}D_m\vett{E}^{(m)}(\vett{r},t),\\
\vett{B}(\vett{r},t) & =\sum_{m\in\mathcal{M}}D_m\vett{B}^{(m)}(\vett{r},t).
\end{align}
\eseq
\subsection{The Role of Polarisation of Hertz Potential in Determining the Properties of Vector RSABs}
Vector beams are frequently used as a model to describe focused light. From this perspective, the method of Hertz potential offers an intuitive and insightful perspective on the process of focussing of a beam of light by a lens, or an objective, for example. In fact, one can interpret the Hertz potential 
$\boldsymbol\Pi$ as the electromagnetic field before the focussing system, consisting of a scalar field distribution, and a given polarisation $\uvett{f}$. The vectorialisation procedure described in Eqs. \eqref{eq5}, then, represents the full vector field after the focussing process (for example, in the focal plane of a lens). Because of the structure of Eqs. \eqref{eq5}, it is not difficult to see, that the initial polarisation $\uvett{f}$ possessed by the field will contribute in determining all the components of the focussed field. 

For the case of RSABs, it is interesting to see whether the vectorialisation procedure described above (i.e., the focussing process) preserves their self-accelerating character, or, in case it does not, under which conditions the self-accelerating character of RSABs is preserved. To do so, first we introduce the polarisation vector $\uvett{f}=f_p\uvett{x}+f_s\uvett{y}$ (where $f_{p,s}\in\mathbb{C}$, and $|f_p|^2+|f_s|^2=1$). Then, we use Eqs. \eqref{eq6bis} and \eqref{eq8} to calculate the vector electric and magnetic fields corresponding to arbitrary polarised Bessel beams. Because of the intrinsic cylindrical symmetry of RSABs, we also introduce a (normalised) cylindrical reference frame $\{\uvettGreek{\rho},\uvettGreek{\theta},\uvettGreek{\zeta}\}$. In this reference frame, the electric and magnetic fields of a single vector Bessel component can be written as 
\bseq\label{eq10}
\begin{align}
\vett{E}^{(m)}(\vettGreek{\rho},t) &= e^{i(\beta\zeta-\omega t+m\Phi)}\left[E^{(m)}_{\rho}(\vettGreek{\rho})\uvettGreek{\rho}+E^{(m)}_{\theta}(\vettGreek{\rho})\uvettGreek{\theta}+E^{(m)}_{\zeta}(\vettGreek{\rho})\uvettGreek{\zeta}\right],\\
\vett{B}^{(m)}(\vettGreek{\rho},t) &= e^{i(\beta\zeta-\omega t+m\Phi)}\left[B^{(m)}_{\rho}(\vettGreek{\rho})\uvettGreek{\rho}+B^{(m)}_{\theta}(\vettGreek{\rho})\uvettGreek{\theta}+ B^{(m)}_{\zeta}(\vettGreek{\rho})\uvettGreek{\zeta}\right],\\
\end{align}
\eseq
where $\Phi=\theta+\Lambda\zeta$ is the co-rotating coordinate defined in the previous section, and the field components are given by
\bseq\label{eq11}
\begin{align}
E^{(m)}_{\rho}(\vettGreek{\rho})&=\omega(\beta+m\Lambda)(f_s\cos\theta-f_p\sin\theta)\text{J}_m(\alpha_m\rho),\\
E^{(m)}_{\theta}(\vettGreek{\rho})&=-\omega(\beta+m\Lambda)(f_p\cos\theta+f_s\sin\theta)\text{J}_m(\alpha_m\rho),\\
E^{(m)}_{\zeta}(\vettGreek{\rho})&=\frac{m\omega}{\rho}\text{J}_m(\alpha_m\rho)(f_p\cos\theta+f_s\sin\theta)\nonumber\\
&+i\omega\left[\alpha_m\text{J}_{m-1}(\alpha_m\rho)-\frac{m}{\rho}\text{J}_{m}(\alpha_m\rho)\right](f_s\cos\theta-f_p\sin\theta),
\end{align}
\eseq
for the electric field, and
\bseq\label{eq12}
\begin{align}
B^{(m)}_{\rho}(\vettGreek{\rho})&=-\frac{\alpha_m}{\rho}\left[\left(f_p-imf_s\right)\cos\theta+\left(f_s+imf_p\right)\sin\theta\right]\text{J}_{m}^{'}(\alpha_m\rho)\nonumber\\
&-2im\text{J}_m(\alpha_m\rho)\left[\left(f_s+imf_p\right)\cos\theta-\left(f_p-imf_s\right)\sin\theta\right],\\
B^{(m)}_{\theta}(\vettGreek{\rho})&=\frac{im}{\rho}\left(f_p\cos\theta+f_s\sin\theta\right)\left[\alpha_m\text{J}_m^{'}(\alpha_m\rho)-\text{J}_m(\alpha_m\rho)\right]\nonumber\\
&+\left(f_s\cos\theta-f_p\sin\theta\right)\left[(\beta+m\Lambda)^2\text{J}_m(\alpha_m\rho)-\alpha^2\text{J}_m^{''}(\alpha_m\rho)\right],\\
B^{(m)}_{\zeta}(\vettGreek{\rho})&=-\frac{m(\beta+m\Lambda)}{2}\left(f_s\cos\theta-f_p\sin\theta\right)\text{J}_m(\alpha_m\rho)\nonumber\\
&+i\alpha_m(\beta+m\Lambda)\left(f_p\cos\theta+f_s\sin\theta\right)\text{J}_m^{'}(\alpha_m\rho),
\end{align}
\eseq
for the magnetic field. In the equations above, $\text{J}_{m}^{'}(\alpha_m\rho)$, and $\text{J}_{m}^{''}(\alpha_m\rho)$  are the first and second derivative of the Bessel function with respect to their argument, respectively \cite{nist}. The electric and magnetic fields of arbitrary polarised RSABs can be then constructed by substituting the expresisons above into Eqs. \eqref{eq11}. Their explicit expression is reported in Appendix A, for completeness.
\subsection{Polarisation Constraint for Vector RSABs}
In Sect. 2, we have described the requirements that a scalar field must fulfill, in order to be a RSAB. In particular, the most important requirement is the existence of a suitable co-rotating reference frame, in which the field appears propagation invariant. If such reference frame exists, an observer at rest in such reference frame would then experience a fictitious centrifugal force. 

For scalar fields, however, this condition is independent on polarisation, as it only applies to the field distribution, and not to the constant polarisation pattern possessed by the field. For vector beams, on the other hand, this assumption may not be valid anymore, as different polarisation states are focussed in different ways, thus resulting in a mixing of the various field components \cite{bornWolf}. In this case, then, it is necessary to investigate under which condition the polarisation coefficients $f_p$ and $f_s$ preserve the self-accelerating character of vector RSABs. A natural way to prove this, is to impose that vector RSABs fulfill the same requirements described in Sect. 2. 

To do so, we first define a suitable co-rotating reference frame, in which the electric and magnetic fields of a vector RSAB appear propagation invariant. If such reference frame exists, this automatically implies that the self-accelerating character has been preserved by the vectorialisation procedure. This means, that the electric (magnetic) field described by the first (second) of Eqs. \eqref{eq9} must be propagation invariant in a co-rotating reference frame $\vettGreek{\rho}'=\mathcal{S}\vettGreek{\rho}$ defined by  the following coordinate transformation
\beq
\left\{\begin{array}{ll}
\rho'=\rho,\\
\Phi=\theta+\Lambda\zeta,\\
\zeta'=\zeta.
\end{array}\right.
\eeq
In principle, one should check that both the electric and the magnetic field are independently propagation invariant in this reference frame. However, Maxwell's equation impose that if one field fulfills the requirement, the other must fulfill it too. For this reason, we limit or analysis to the electric field only. The same condition that we will derive for the polarisation coefficients $f_p$ and $f_s$ will apply to the magnetic field as well, and can be also derived using the same approach with the magnetic, rather than electric, field. 

We then start by separating the electric field into its transverse and longitudinal parts, namely  $\vett{E}(\vettGreek{\rho}')=\vett{E}_{\perp}(\vettGreek{\rho}')+\vett{E}_{\parallel}(\vettGreek{\rho}')$, and require that they are both propagation invariant, i.e., $\partial\vett{E}_{\perp,\parallel}(\vettGreek{\rho}')/\partial\zeta=0$. Instead of dealing directly with this condition, however, we can require that the transverse and longitudinal intensities, rather than amplitudes, are propagation invariant. By doing this, we are formally requiring that only intensity rotating RSABs remain propagation invariant upon focussing. However, if the intensity of a field is independent from $\zeta$, its amplitude will be $\zeta$-independent as well, and the $\zeta$ dependence can be at most contained into a phase factor. Once the condition on the intensity has been met, one could then look at the phase of the corresponding field, and check, whether it remains synchronised with its corresponding intensity profile.

The transverse  $\left|\vett{E}_{\perp}(\vettGreek{\rho})\right|^2=\left|E_{\rho}\right|^2+\left|E_{\theta}\right|^2$, and longitudinal $\left|\vett{E}_{\parallel}(\vettGreek{\rho})\right|^2=\left|E_{\zeta}\right|^2$ intensities can be calculated using the expressions given in Appendix A, thus obtaining
\bseq\label{eq14}
\begin{align}
\left|\vett{E}_{\perp}(\vettGreek{\rho})\right|^2 &= \sum_{m\in\mathcal{M}}\left|\mathcal{E}_m^{(1)}(\rho)\right|^2+2\sum_{n\neq m\in\mathcal{M}}\mathcal{E}_m^{(1)}(\rho)\left[\mathcal{E}_n^{(1)}(\rho)\right]^*\cos\left[\left(m-n\right)\Phi\right],\\
\left|\vett{E}_{\parallel}(\vettGreek{\rho})\right|^2 &=G_1(\rho,\theta)+2\sum_{n\neq m\in\mathcal{M}}\mathcal{E}_m^{(2)}(\rho)\left[\mathcal{E}_n^{(3)}(\rho)\right]^*\Big\{a_pa_s\cos^2\theta\sin\left[\left(m-n\right)\Phi-\Delta\right]\nonumber\\
&-(a_p^2-a_s^2)\sin\theta\cos\theta\sin\left[\left(m-n\right)\Phi\right]-a_pa_s\sin^2\theta\sin\left[\left(m-n\right)\Phi+\Delta\right]\Big\}\label{parallel}
\end{align}
\eseq
where we have rewritten the polarisation coefficients as $f_p=a_p$, $f_s=a_s\exp{(i\Delta)}$ (with $a_p, a_s, \Delta \in \mathbb{R}$), being $\Delta$ the relative phase between the two polarisation components,  and
\begin{eqnarray}\label{G1}
G_1(\rho,\theta)&=&\sum_{m\in\mathcal{M}}\Big\{a_p^2\left[\cos^2\theta \left|\mathcal{E}_m^{(2)}(\rho)\right|^2+\sin^2\theta \left|\mathcal{E}_m^{(3)}(\rho)\right|^2\right]\nonumber\\
&+&a_s^2\left[\cos^2\theta \left|\mathcal{E}_m^{(3)}(\rho)\right|^2+\sin^2\theta \left|\mathcal{E}_m^{(2)}(\rho)\right|^2\right]\nonumber\\
&+&2a_pa_s\left(\left|\mathcal{E}_m^{(2)}(\rho)\right|^2-\left|\mathcal{E}_m^{(3)}(\rho)\right|^2\right)\sin\theta\cos\theta\cos\Delta\Big\}.
\end{eqnarray}
Equations \eqref{eq14} already contain an important information. No matter the polarisastion, the transverse intensity always remains propagation invariant, as no explicit $\zeta$-dependence appears in the expression of $\left|\vett{E}_{\perp}(\vettGreek{\rho})\right|^2$. 

The longitudinal part of the intensity, on the other hand, 
contains terms that depend on $\sin\theta$ and $\cos\theta$. Once transformed in the co-rotating frame, these terms become $\zeta$-dependent, as $\theta=\Phi-\Lambda\zeta$. To avoid this problem, the polarisation coefficients must be chosen in such a way to guarantee the propagation invariance of the longitudinal intensity as well.  The condition on $a_p$, $a_s$, and $\Delta$ can be then found by requiring that 
\barr\label{eq15}
\frac{\partial\left|\vett{E}_{\parallel}(\vettGreek{\rho})\right|^2}{\partial\zeta} &=\left(a_p^2-a_s^2\right)\Big\{\left[F_2(\rho)-F_3(\rho)\right]\sin2\theta+F_4(\rho)\cos2\theta\Big\}\nonumber\\
&+a_pa_s\cos\Delta\left\{F_4(\rho)\sin2\theta-\left[F_2(\rho)-F_3(\rho)\right]\cos2\theta\right\}=0,
\earr
where the functions $F_k(\rho)$ (with $k=\{1,2,3,4\}$) can be determined from Eqs. \eqref{parallel} and \eqref{G1}. It is not difficult to see, that the above equation is satisfied if and only if $a_p=a_s$, and $\Delta=\pm\pi/2$. Moreover, since $|f_p|^2+|f_s|^2=1$, this condition implies that $f_p=1/\sqrt{2}$, and $f_s=\pm i/\sqrt{2}$., which correspond to left-handed ($+$) and right-handed ($-$) circular polarisation, respectively

This is the main result of our work. Vector RSABs only maintain their self-accelerating character if the polarisation of the Hertz vector is chosen to be circular. In other words, when focussing polarised RSABs, only circular polarisation is allowed, in order to preserve the self-accelerating character of the focused RSABs.  

A simple explanation of this result can be given by looking at the symmetry of the scalar and vector beams, respectively. In the scalar case, in fact, RSABs naturally possess cylindrical symmetry, due to their transverse profile. By virtue of this symmetry, the co-rotating coordinate can be chosen as a $\zeta$-dependent azimuthal coordinate, namely $\Phi=\theta+\Lambda\zeta$. Upon focussing, the overall cylindrical symmetry must be preserved, in order for the vector RSAB to maintain its self-accelerating character. This, ultimately, constraints the polarisation to be chosen as circular.
\subsection{Vector Fields from Circularly Polarised RSABs}
We now apply the polarisation constraints derived above and investigate the form of the electric and magnetic fields generated by focussing circularly polarised RSABs. By substituting $f_p=1/\sqrt{2}$ and $f_s=i\sigma/\sqrt{2}$ into Eqs. \eqref{eq11} and \eqref{eq12}, and using Eqs. \eqref{eq9}, the electric and magnetic fields of a circularly polarised focussed RSAB can be written as
\bseq\label{eq16}
\begin{align}
\vett{E}(\vettGreek{\rho},t)&=\sum_{m\in\mathcal{M}}\vett{e}_m(\rho)\,e^{i[(m+\sigma)\theta+(\beta+m\Lambda)\zeta-\omega t]},\\
\vett{B}(\vettGreek{\rho},t)&=\sum_{m\in\mathcal{M}}\vett{b}_m(\rho)\,e^{i[(m+\sigma)\theta+(\beta+m\Lambda)\zeta-\omega t]},
\end{align}
\eseq
where $\sigma=\pm 1$ is the helicity index, which distinguishes between left-handed ($+$) and right-handed ($-$) circular polarisation \cite{mandelWolf}, $\vett{e}_m(\rho)$, and $\vett{b}_m(\rho)$ are  radially dependent vector field, whose explicit expression, is given by
\bseq\label{eq16bis}
\begin{align}
\vett{e}_m(\vettGreek{\rho})&=\sigma D_m\omega\left\{ i(\beta+m\Lambda)\text{J}_m(\alpha_m\rho)\,\uvett{h}_{\sigma}-\left[\alpha_m\text{J}_{m-1}(\alpha_m\rho)-\frac{m}{\rho}\text{J}_m(\alpha_m\rho)\right]\uvettGreek{\zeta}\right\},\\
\vett{b}_m(\rho) &= \frac{D_m}{\sqrt{2}}\Big\{(1+m\sigma)\left[\frac{\alpha_m}{\rho}\text{J}_m^{'}(\alpha_m\rho)+2m\sigma\text{J}_m(\alpha_m\rho)\right]\uvettGreek{\rho}\nonumber\\
&+i\left\{\frac{m}{\rho}\left[\text{J}_m^{'}(\alpha_m\rho)-\text{J}_m(\alpha_m\rho)\right]+\sigma\left[(\beta+m\Lambda)^2\text{J}_m(\alpha_m\rho)-\alpha_m^2\text{J}_m^{''}(\alpha_m\rho)\right]\right\}\uvettGreek{\theta}\nonumber\\
&+i(\beta+m\Lambda)\left[\alpha_m\text{J}_m^{'}(\alpha_m\rho)-\frac{m\sigma}{2}\text{J}_m(\alpha_m\rho)\right]\uvettGreek{\zeta}\Big\},
\end{align}
\eseq
where $\uvett{h}_{\sigma}=\left(\uvettGreek{\rho}+i\sigma\uvettGreek{\theta}\right)/\sqrt{2}=\left(\uvett{x}+i\sigma\uvett{y}\right)\sqrt{2}$ is the helicity basis \cite{mandelWolf}.
\begin{figure}[!t]
\begin{center}
\includegraphics[width=\textwidth]{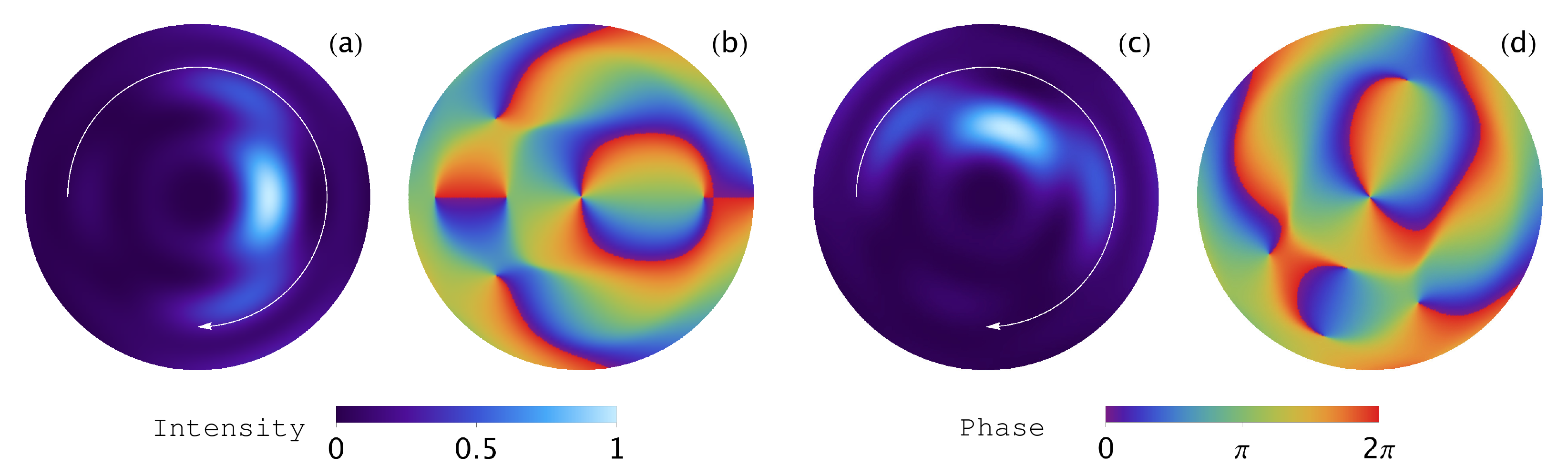}
\caption{Intensity and phase distribution for the longitudinal component $E_{\zeta}$ of the electric field described by Eq. \eqref{eq17a}, for $\sigma=1$.  Panels (a) and (b) correspond to the intensity and phase distributions at $z=0$, while panels (c), (d) to $z=0.8(2\pi/\Lambda)$. These plots are made assuming $0\leq\rho\leq 10$.  The plot parameters are the same as the one chosen for Fig. \ref{figure1}. The white arrow in the intensity profiles show the direction of rotation.}
\label{figure2}
\end{center}
\end{figure}
For experimentally realisable RSABs, $\Lambda\ll 1$. Within this approximation, one should distinguish between field rotating, and intensity rotating vector RSABs. For the former, $\beta=0$, and the radial and azimuthal components of the electric field, as well as the longitudinal component of the magnetic field, are $\mathcal{O}(\Lambda)$, and can therefore be neglected, leaving a purely longitudinal electric field, and a purely transverse magnetic field, namely
\bseq\label{eq17}
\begin{align}
\vett{E}(\vettGreek{\rho},t) &\simeq\left(-\frac{\sigma\omega}{\sqrt{2}}\right)\sum_{m\in\mathcal{M}}D_m\left[\alpha_m\text{J}_{m-1}(\alpha_m\rho)-\frac{m}{\rho}\text{J}_m(\alpha_m\rho)\right]e^{i[(m+\sigma)\theta+m\Lambda\zeta-\omega t]}\uvettGreek{\zeta},\label{eq17a}\\
\vett{B}(\vettGreek{\rho},t)&\simeq\frac{1}{\sqrt{2}}\sum_{m\in\mathcal{M}}D_m\Bigg\{(1+m\sigma)\left[\frac{\alpha_m}{\rho}\text{J}_m^{'}(\alpha_m\rho)+2m\sigma\text{J}_m(\alpha_m\rho)\right]\uvettGreek{\rho}\nonumber\\
&+i\left[\frac{m}{\rho}\text{J}_m^{'}(\alpha_m\rho)-\frac{m}{\rho}\text{J}_m(\alpha_m\rho)-\sigma\alpha_m^2\text{J}_m^{''}(\alpha_m\rho)\right]\uvettGreek{\theta}\Bigg\}e^{i[(m+\sigma)\theta+m\Lambda\zeta-\omega t]}.
\end{align}
\eseq
For intensity rotating vector RSABs, and within the paraxial approximation, $(\beta+m\Lambda)\simeq 1$, and therefore $\alpha_m\ll 1$. In this case, all three components  of the electric and magnetic field are nonzero, and assume the following, simplified, form:
\bseq\label{eq19}
\begin{align}
\vett{E}(\vettGreek{\rho},t) &\simeq\sum_{m\in\mathcal{M}}\sigma\omega D_m\left\{i\text{J}_m(\alpha_m\rho)\,\uvett{h}_{\sigma}-\frac{m}{\rho}\text{J}_m(\alpha_m\rho)\uvettGreek{\zeta}\right\}e^{i[(m+\sigma)\theta+(\beta+m\Lambda)\zeta-\omega t]}\label{eq19a},\\
\vett{B}(\vettGreek{\rho},t)&\simeq\sum_{m\in\mathcal{M}}\frac{D_m}{\sqrt{2}}\Bigg\{2m(\sigma+m)\text{J}_m(\alpha_m\rho)\uvettGreek{\rho}+i\left[\frac{m}{\rho}\text{J}_m^{'}(\alpha_m\rho)-\frac{m}{\rho}\text{J}_m(\alpha_m\rho)\right]\uvettGreek{\theta}\nonumber\\
&-\frac{i m\sigma}{2}\text{J}_m(\alpha_m\rho)\uvettGreek{\zeta}\Bigg\}e^{i[(m+\sigma)\theta+(\beta+m\Lambda)\zeta-\omega t]}.
\end{align}
\eseq
From the expressions above, it appears clear that, upon focussing, the electric field maintains the original circular polarisation (in the transverse plane) of the focussed beam, while the polarisation of the magnetic field gets mixed. This, however, is only a result of the fact that we only considered TE fields to start with. If one would repeat the above calculations for TM fields, in fact, the result would be the same, with the magnetic field retaining the original polarisation and the electric field being mixed up. In the most general case, where both TE and TM waves are present, each field has these two components of polarisation, thus resulting in a more complex polarisation pattern. 

\begin{figure}[!t]
\begin{center}
\includegraphics[width=\textwidth]{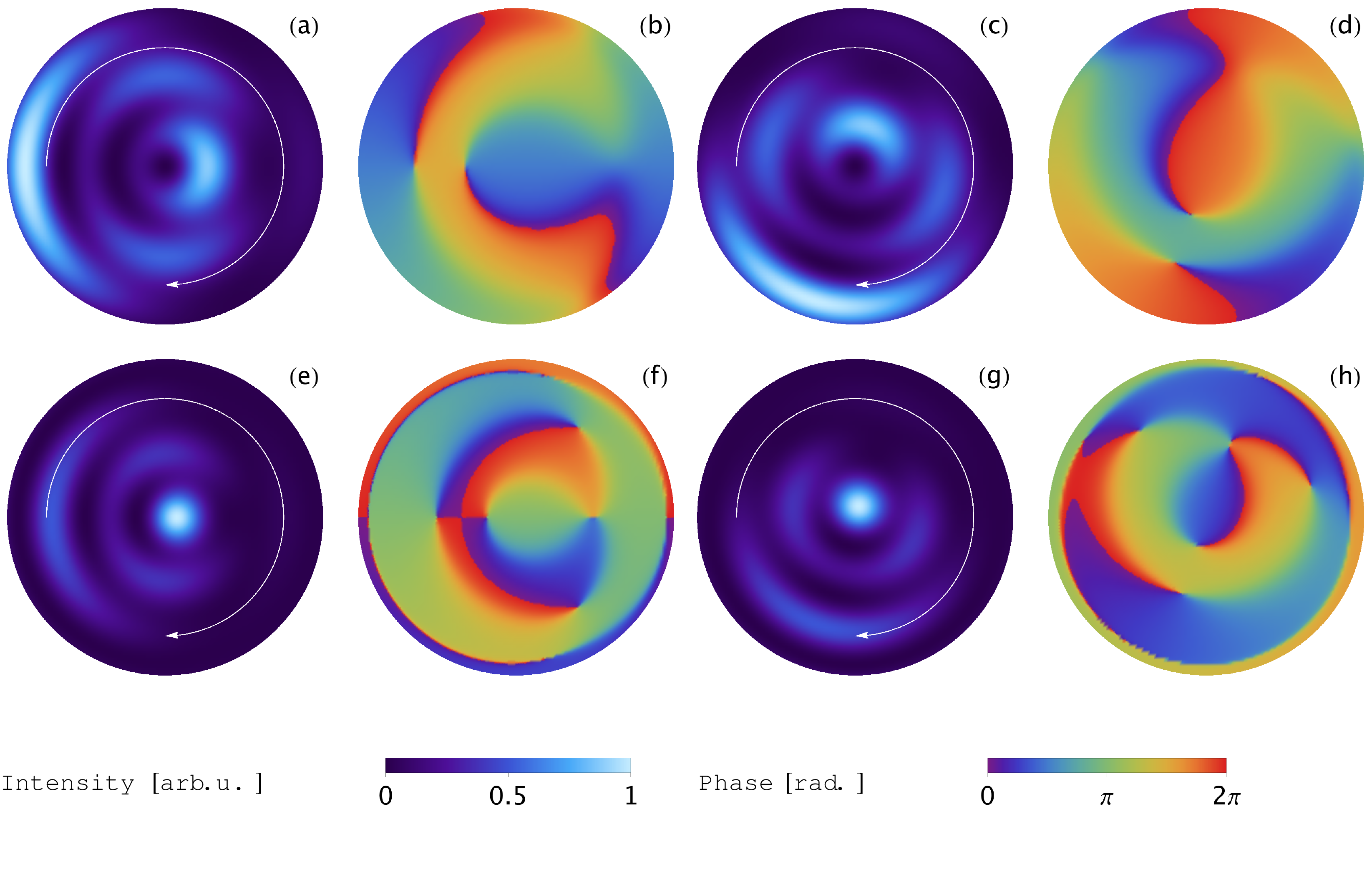}
\caption{Intensity and phase distribution for the radial (top row) and longitudinal (bottom row) components  of the electric field described by Eq. \eqref{eq19a}, for $\sigma=1$.  Panels (a), (b), (e), and (f) correspond to the intensity and phase distributions at $z=0$, while panels (c), (d), (g), and (h) to $z=0.8(2\pi/\Lambda)$. These plots are made assuming $0\leq\rho\leq 1200$.  The azimuthal component of the field is not shown, as its intensity profile is the same as the radial one [ see Eq. \eqref{eq19a}]. The plot parameters are the same as the one chosen for Fig. \ref{figure1}. The white arrow in the intensity profiles show the direction of rotation.}
\label{figure3}
\end{center}
\end{figure}

The intensities and phases of the electric field components for field rotating and intensity rotating RSABs are reported in Figs. \ref{figure2}, and \ref{figure3}, respectively.
As it can be seen from Fig. \ref{figure2}(d), upon focussing, field rotating vector RSABs lose their property, that intensity and phase profile are synchronised in rotation during propagation. This, ultimately is due to the fact that while the field intensity contains terms of the form $\cos\left[\left(m-n\right)\left(\theta+\Lambda\zeta\right)\right]$, the phase contains terms that oscillate like $\cos\left[m\left(\theta+\Lambda\zeta\right)+\sigma\theta\right]$. The presence of the extra term $\sigma\theta$ (which disappears in the intensity) is then responsible for the different evolution of amplitude and phase of the field, as it corresponds to a $\zeta$-dependent term, once transformed in the co-rotating frame.
\section{Linear and Angular Momentum Densities of Vector RSABs}
In this section, we calculate the linear and angular momentum for intensity rotating vector RSABs. We limit ourselves to the paraxial case, as within this approximation, we can separate the angular momentum in its spin and orbital parts. This gives us the possibility to distinguish between intrinsic and extrinsic orbital angular momentum of vector RSABs, and to then isolate the extrinsic contribution given by the fact that the intensity rotates with angular velocity $\Lambda$. 

Following Jackson, the linear and angular momentum of the electromagnetic field are defined as follows \cite{jackson}:
\bseq\label{eq21}
\begin{align}
\vett{P}=\int\,d^2\rho\,\vett{p}(\vettGreek{\rho}),\\
\vett{J}=\int\,d^2\rho\,\vett{j}(\vettGreek{\rho}),
\end{align}
\eseq
where
\bseq\label{eq20}
\begin{align}
\vett{p}(\vettGreek{\rho})=\frac{\varepsilon_0}{2}\,\operatorname{Re}\left\{\vett{E}(\vettGreek{\rho})\times\vett{B}^*(\vettGreek{\rho})\right\},\label{eq20a}\\
\vett{j}(\vettGreek{\rho})=\frac{\varepsilon_0}{2}\,\operatorname{Re}\left\{\vettGreek{\rho}\times\vett{p}(\vettGreek{\rho})\right\},\label{eq20b}
\end{align}
\eseq
are the correspondent densities, $d^2\rho=\rho d\rho d\theta$, and the integrals are extended over the whole space. 

As it can be seen from Eqs. \eqref{eq3} and \eqref{eq16}, RSABs are defined in terms of superpositions of Bessel beams. Therefore, as Bessel beams cary infinite energy, the above integrals diverge, and linear and angular momentum (as well as energy) are not well defined quantities for RSABs. This problem, however, can be overcome in different ways, by introducing different forms of regularisation. For example, one could limit the radial integration, up to a maximum radius. Alternatively, one could insert a regularisation function, such a Gaussian function, in the radial integrals to make them finite. Physically speaking, both regularisations can be implemented. The former, in fact, corresponds to use a pupil of a fixed diameter to filter the field. The latter, on the other hand, corresponds to describe RSABs in terms of Bessel-Gauss beams, which, de facto, are the closest approximation to Bessel beams that can be realised experimentally.

In the remaining of this section, we calculate the explicit expressions for both the momentum densities, and their integrated counterpart. For the sake of simplicity, however, we will not compute the radial integrals. These, in fact, only contribute to a multiplicative constant, and do not carry any valuable information for the purpose of investigating the properties of linear and angular momentum of RSABs.
\subsection{Linear Momentum}
\begin{figure}[!t]
\begin{center}
\includegraphics[width=\textwidth]{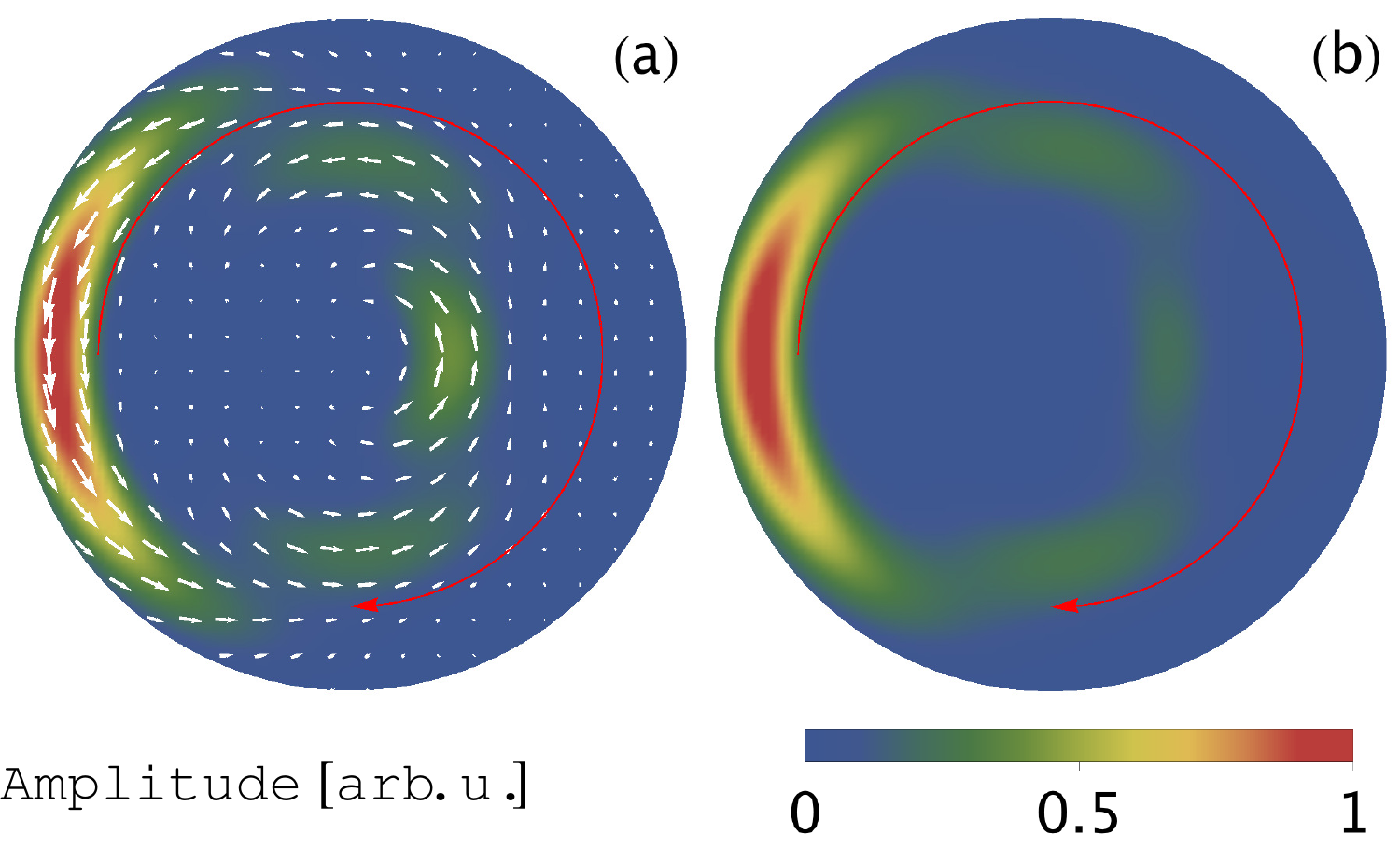}
\caption{Transverse (a) and longitudinal (b) components of the linear momentum density, as given by Eq. \eqref{eq22}, in the plane $\zeta=0$, for $\sigma=1$. The white arrows in panel (a) represent the flow of the transverse component of the linear momentum density. As it can be seen, the transverse momentum density always points in the opposite direction with respect to the field rotation (red arrow). These plots are made assuming $0\leq\rho\leq 1200$. The plot parameters are the same as the one chosen for Fig. \ref{figure1}. The red arrow in both panels show the direction of rotation of the RSAB intensity.}
\label{figure5}
\end{center}
\end{figure}

If we substitute the expressions of the electric and magnetic fields of a paraxial RSABs as given by Eqs. \eqref{eq16} into Eq. \eqref{eq20a}, the linear momentum density can be written as follows:
\barr\label{eq22}
\vett{p}(\vettGreek{\rho}) &=& \sum_{m,n\in{\mathcal{M}}}\frac{|D_mD_n|\omega\varepsilon_0}{4}\Big\{P_{\rho}^{(m,n)}(\rho)\sin\left[\left(m-n\right)\left(\theta+\Lambda\zeta\right)+\phi_m-\phi_n\right]\uvettGreek{\rho}\nonumber\\
&+&P_{\theta}^{(m,n)}(\rho)\cos\left[\left(m-n\right)\left(\theta+\Lambda\zeta\right)+\phi_m-\phi_n\right]\uvettGreek{\theta}\nonumber\\
&+&P_{\zeta}^{(m,n)}(\rho)\cos\left[\left(m-n\right)\left(\theta+\Lambda\zeta\right)+\phi_m-\phi_n\right]\uvettGreek{\zeta}\Big\},
\earr
where $\phi_{m,n}=\arg[D_{m,n}]$, and
\bseq\label{eq23}
\begin{align}
P_{\rho}^{(m,n)}(\rho) &=\sigma\text{J}_m(\alpha_m\rho)\Bigg\{\left[\frac{mn}{\rho^2}-\frac{m\sigma(\beta+n\Lambda)^2}{\rho}-\frac{n(\beta+m\Lambda)(\beta+n\Lambda)}{2}\right]\text{J}_n(\alpha_n\rho)\nonumber\\
&-\frac{mn}{\rho^2}\text{J}_n^{'}(\alpha_n\rho)\Bigg\},\\
P_{\theta}^{(m,n)}(\rho) &=\sigma\left[\frac{2mn(n+\sigma)}{\rho}+\frac{n\sigma(\beta+m\Lambda)(\beta+n\Lambda)}{2}\right]\text{J}_m(\alpha_m\rho)\text{J}_n(\alpha_n\rho),\\
P_{\zeta}^{(m,n)}(\rho) &=(\beta+m\Lambda)\text{J}_m(\alpha_m\rho)\Bigg\{\left[(\beta+n\Lambda)^2+2n(n+\sigma)-\frac{n\sigma}{\rho}\right]\text{J}_n(\alpha_n\rho)\nonumber\\
&+\frac{n}{\rho}\text{J}_n^{'}(\alpha_n\rho)\Bigg\},
\end{align}
\eseq
where terms of order $\mathcal{O}(\alpha_m)$ have been neglected, since in the paraxial regime $\alpha_m\ll 1$.  The components of the linear momentum density are shown in Fig. \ref{figure5}. Notice, that the transverse part of the linear momentum presents an unusual characteristic. While it rotates clockwise along the propagation direction, as the intensity distribution of the correspondent RSAB does, the local orientation of the transverse momentum is purely azimuthal (despite $\vett{p}(\vettGreek{\rho})$ has a nonzero radial component), and always directed in the opposite direction, with respect to the rotation direction of the RSAB, as it can be seen from the white arrows in Fig. \ref{figure5}(a). This has an interesting consequence for applications such particle manipulation and material processing, where the local, rather than the global, behaviour of the momentum plays an important role. While the RSAB (and, with it, the transverse momentum density) rotates clockwise during propagation, a particle placed in the vicinity of a RSAB will experience a local momentum, that will tend to push it in the opposite direction. This effect, however, is purely local, and it disappears when considering the whole momentum. 
To understand this, let us integrate Eq. \eqref{eq22} over the transverse space. The linear momentum can be then written as follows
\beq\label{momentumP}
\vett{P}=\sum_{m\in\mathcal{M}}|D_m|^2\Bigg[\mathcal{P}^{(m)}_{\theta}\uvettGreek{\theta}+\mathcal{P}^{(m)}_{\zeta}\uvettGreek{\zeta}\Bigg],
\eeq
where
\beq\label{Plambda}
\mathcal{P}_{\lambda}^{(m)}=\frac{\pi\omega\varepsilon_0}{2}\int_0^{\infty}\,d\rho\,\rho\,P_{\lambda}^{(m,m)}(\rho),
\eeq
where $\lambda\in\{\theta,\zeta\}$. Notice that the radial integrals (once regularised) amount to a positive constant. Moreover, there is no radial component of the momentum, since the radial part of $\vett{p}(\vett{\rho})$ depends on $\sin[(m-n)(\theta+\Lambda\zeta)+\phi_m-\phi_n]$, which gives zero once integrated with respect to the azimuthal coordinate $\theta$. 
\subsection{Spin and Orbital Angular Momentum}
To calculate the spin and orbital angular momentum for intensity rotating, paraxial RSABs, we make use of the usual decomposition of the total angular momentum in its spin (SAM) and orbital (OAM) components, namely $\vett{J}=\vett{S}+\vett{L}$ \cite{libroOAM}. To do so, we first need to introduce the vector potential $\vett{A}(\vettGreek{\rho})$ associated to the electric and magnetic fields defined above, since the decomposition assumes a rather simple form if expressed in terms of the vector potential. Looking at Eqs. \eqref{eq5}, it is not difficult to see that $\vett{A}=\nabla\times\vettGreek{\Pi}(\vettGreek{\rho},t)$. The explicit expression of $\vett{A}$ for an intensity rotating RSAB is given in Appendix B. 

\begin{figure}[!t]
\begin{center}
\includegraphics[width=\textwidth]{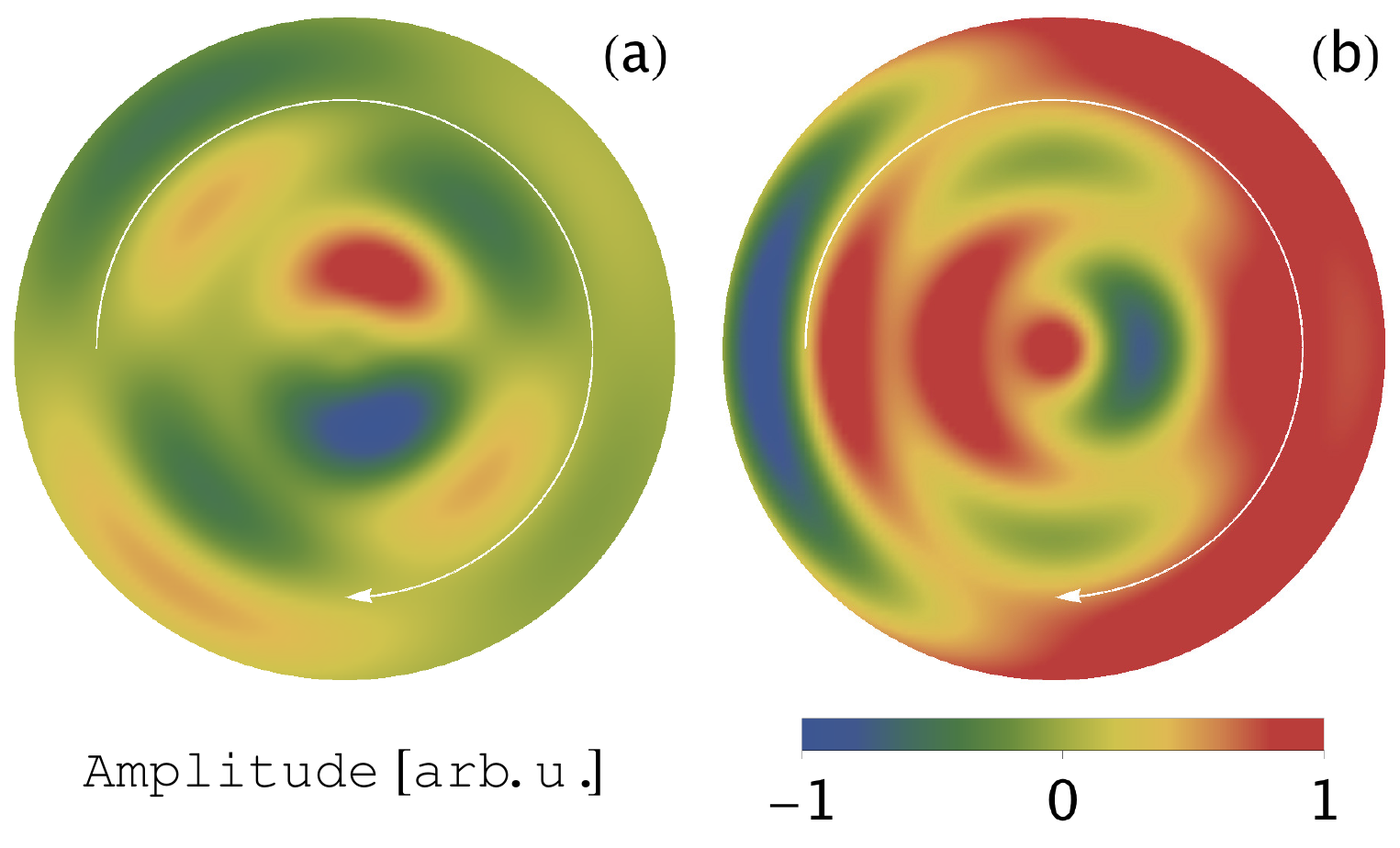}
\caption{Radial (a) and longitudinal (b) components of the SAM density, as given by Eqs. \eqref{SAMdensity}, in the plane $\zeta=0$, for $\sigma=1$. The blue regions in both panels indicate areas of negative SAM density, where the helicity is oriented in the opposite direction, with respect to the propagation direction. The azimuthal component of the SAM density is not reported here, as it is, up to a constant, the same as the radial one. These plots are made assuming $0\leq\rho\leq 1200$. The plot parameters are the same as the one chosen for Fig. \ref{figure1}. The red arrow in both panels show the direction of rotation of the RSAB intensity.}
\label{figure6}
\end{center}
\end{figure}

Following Ref. \cite{libroOAM}, the angular momentum density then assumes the following form
\beq\label{eq24}
\mathbf{j}(\vettGreek{\rho})=\vett{s}(\vettGreek{\rho})+\vett{l}(\vettGreek{\rho})=\frac{\varepsilon_0}{2}\operatorname{Re}\left\{-i\vett{A}^*\times\vett{A}\right\}+\frac{\varepsilon_0}{2}\operatorname{Re}\left\{\vett{A}^*\cdot\left(-i\vettGreek{\rho}\times\nabla\right)\vett{A}\right\},
\eeq
where $-i\vettGreek{\rho}\times\nabla$ is the angular momentum operator \cite{jackson} in the normalised cylindrical reference frame $\{\uvettGreek{\rho},\uvettGreek{\theta},\uvettGreek{\zeta}\}$. Using the expression for the vector potential given in Appendix B, the SAM and OAM of a paraxial, intensity rotating vector RSAB are given as follows:
\bseq\label{eq25}
\begin{align}
\vett{s}(\vettGreek{\rho})&=\frac{\varepsilon_0}{2}\sum_{m,n\in\mathcal{M}}\Bigg\{-S_{\rho}^{(m,n)}\sin\left[\left(m-n\right)\left(\theta+\Lambda\zeta\right)+\phi_m-\phi_n\right]\,\uvettGreek{\rho}\nonumber\\
&+S_{\theta}^{(m,n)}\cos\left[\left(m-n\right)\left(\theta+\Lambda\zeta\right)+\phi_m-\phi_n\right]\,\uvettGreek{\theta}\nonumber\\
&-S_{\zeta}^{(m,n)}(\rho)\cos\left[\left(m-n\right)\left(\theta+\Lambda\zeta\right)+\phi_m-\phi_n\right]\,\uvettGreek{\zeta}\Bigg\},\\
\vett{l}(\vettGreek{\rho}) &=\frac{\varepsilon_0}{2}\sum_{m,n\in\mathcal{M}}\Bigg\{L_{\rho}^{(m,n)}(\rho)\sin\left[\left(m-n\right)\left(\theta+\Lambda\zeta\right)+\phi_m-\phi_n\right]\,\uvettGreek{\rho}\nonumber\\
&-L_{\theta}^{(m,n)}(\rho)\cos\left[\left(m-n\right)\left(\theta+\Lambda\zeta\right)+\phi_m-\phi_n\right]\,\uvettGreek{\theta}\nonumber\\
&+L_{\zeta}^{(m,n)}(\rho)\cos\left[\left(m-n\right)\left(\theta+\Lambda\zeta\right)+\phi_m-\phi_n\right]\,\uvettGreek{\zeta},
\end{align}
\eseq
where
\barr\label{SAMdensity}
S_{\rho}^{(m,n)}(\rho)&=&\frac{\mathcal{D}_{m,n}(\rho)\beta(m-n)}{\rho},\\
S_{\theta}^{(m,n)}(\rho)&=&\sigma S_{\rho}(\rho),\\
S_{\zeta}^{(m,n)}(\rho)&=&2\sigma\mathcal{D}_{m,n}(\rho)(\beta+m\Lambda)(\beta+n\Lambda),
\earr
are the components of the spin angular momentum density, while
\barr
L_{\rho}^{(m,n)}(\rho)&=&\mathcal{D}_{m,n}(\rho)(\beta+m\Lambda)\left[\sigma\frac{n(m+\sigma)}{\rho}+\rho(\beta+m\Lambda)(\beta+n\Lambda)\right],\\
L_{\theta}^{(m,n)}(\rho)&=&\sigma L_{\rho}(\rho),\\
L_{\zeta}^{(m,n)}(\rho)&=&\mathcal{D}_{m,n}(\rho)(\beta+m\Lambda)\left[\frac{m(\beta+m\Lambda)+n(m+\sigma)}{\rho}\right],
\earr
are the components of the orbital angular momentum density, and $\mathcal{D}_{m,n}(\rho)=|D_mD_n|\text{J}_m(\alpha_m\rho)\text{J}_n(\alpha_n\rho)/2$. In the above expressions, terms of order $\mathcal{O}(\alpha_m)$ have been neglected, since, for paraxial fields, $\alpha_m\ll 1$.  The longitudinal and transverse SAM densities are plotted in Fig. \ref{figure6}. As it can be seen, the SAM density can become negative. This means, that locally, the helicity of the vector RSAB can change sign. However, a close comparison between the SAM density distribution in Fig. \ref{figure6} and the transverse and longitudinal intensity distributions depicted in Fig. \ref{figure3} reveals, that regions of negative SAM density occur where the RSAB intensity is very low, or even zero. 

The spin and orbital angular momenta of paraxial vector RSABs are then obtained by integrating Eqs. \eqref{eq25} over the transverse space. By doing so we obtain
\bseq\label{eq26}
\begin{align}
\vett{S}&=\sigma\sum_{m\in\mathcal{M}}|D_m|^2\mathcal{S}^{(m)}_{\zeta}\uvettGreek{\zeta}\label{eq26a},\\
\vett{L}&=\sum_{m\in\mathcal{M}}|D_m|^2\Bigg[\mathcal{L}^{(m)}_{\theta}\uvettGreek{\theta}+\mathcal{L}^{(m)}_{\zeta}\uvettGreek{\zeta}\Bigg],\label{eq26b}
\end{align}
\eseq
where
\beq
\mathcal{S}_{\zeta}^{(m)}=\frac{\varepsilon_0}{4}\int_0^{\infty}\,d\rho\,\rho\,S_{\zeta}^{(m,m)}(\rho),
\eeq
and
\beq
\mathcal{L}_{\lambda}^{(m)}=\frac{\varepsilon_0}{4}\int_0^{\infty}\,d\rho\,\rho\,L_{\zeta}^{(m,m)}(\rho),
\eeq
being $\lambda\in\{\theta,\zeta\}$. The total SAM is purely longitudinal, and given as the sum of the longitudinal components of the individual Bessel components. This is not surprising, since we are dealing with paraxial fields, for which the SAM is only directed along the propagation direction \cite{libroOAM}.

The OAM, on the other hand, can be seen as the sum of two contributions: an intrinsic component relative to the intrinsic OAM carried by Bessel beams, and an extrinsic one, connected to the fact that the beam rotates around the $\zeta$-axis during propagation. Their explicit expression read then as follows:
\bseq\label{eq27}
\begin{align}
\vett{L}^{(int)}&=\sum_{m\in\mathcal{M}}|\tilde{D}_m|^2\mathcal{L}_{int}^{(m)}\left(\uvettGreek{\theta}+\uvettGreek{\zeta}\right),\\
\vett{L}^{(ext)}&=\sum_{m\in\mathcal{M}}|\tilde{D}_m|^2\left[\sigma\mathcal{L}^{(m)}_{ext,\theta}\uvettGreek{\theta}+\mathcal{L}_{ext,\zeta}^{(m)}\,\uvettGreek{\zeta}\right],
\end{align}
\eseq 
where $\tilde{D}_m=D_m\sqrt{\beta+m\Lambda}$, and
\bseq\label{eq28}
\begin{align}
\mathcal{L}_{int}^{(m)}&=\frac{\varepsilon_0}{4}\, m(m+\sigma)\int_0^{\infty}\,d\rho\,\text{J}_m^2(\alpha_m\rho),\\
\mathcal{L}_{ext,\theta}^{(m)}&=\frac{\varepsilon_0}{4}\,(\beta+m\Lambda)^2\int_0^{\infty}\,d\rho\,\rho^2\text{J}_m^2(\alpha_m\rho),\\
\mathcal{L}_{ext,\zeta}^{(m)}&=\frac{\varepsilon_0}{4}\,m(\beta+m\Lambda)\int_0^{\infty}\,d\rho\,\text{J}_m^2(\alpha_m\rho).
\end{align}
\eseq
The intrinsic part of the OAM has the standard spin-orbit interaction form, through the mixed term $(m+\sigma)$ \cite{libroOAM}. The extrinsic part, on the other hand, depends on $(\beta+m\Lambda)$, which is, essentially, the angular velocity of the beam along the $\zeta$-axis. The beam rotation, moreover, also induces a longitudinal OAM, which is, as well, proportional to the angular velocity $(\beta+m\Lambda)$.
\section{Conclusions}
In this work, we have analysed the properties of vector RSABs, generated by focussing a scalar, polarised RSAB. Using the method of Hertz potentials as a model for the focussing process, we have demonstrated that only circularly polarised scalar RSABs, when focussed, maintain their self-accelerating character. For this case, we have given explicit expressions of the TE vector electric and magnetic fields for both field and intensity rotating RSABs. In particular, we have shown, that the vectorialisation (focussing) process does not allow anymore the amplitude and phase of field rotating RSABs to rotate synchronously during propagation. Within the paraxial approximation, moreover, we have presented the explicit expressions for the linear and angular momentum densities of intensity rotating RSABs. For SAM, in particular, we have shown, that, locally, the SAM density can be negative, thus meaning a local inversion of the helicity axis. Moreover, for the case of OAM, we have distinguished between the intrinsic and extrinsic contributions, and shown how the rotation of the RSAB around the propagation axis is connected with the extrinsic OAM.

Our work represents a useful guideline for investigating experimentally focussed RSABs and their properties. Moreover, the properties highlighted in this work represent a useful toolbox for studying the interaction of RSABs with matter and dielectric particles. In particular, the fact that locally the linear momentum density flows in the opposite direction, with respect to the overall beam rotation during propagation, could open new possibilities for particle manipulation

\section*{Acknowledgements}
The authors wish to thank the Deutsche Forschungsgemeinschaft (grant SZ 276/17-1) for financial support.

\section*{Appendix A: Explicit Form of RSAB Electric and Magnetic Fields}
The vector electric and magnetic fields for single Bessel beams defined in Eqs. \eqref{eq11} and \eqref{eq12} can be used to write the expressions for the RSAB vector electric and magnetic fields explicitly. Substituting these expressions into  Eqs. \eqref{eq9} we then get
\barr\label{electric}
\vett{E}(\vettGreek{\rho}) &=&e^{i(\beta\zeta-\omega t)} \sum_{m\in\mathcal{M}}e^{im\Phi}\Big\{\mathcal{E}_m^{(1)}(\rho)\Big[\left(f_s\cos\theta-f_p\sin\theta\right)\uvettGreek{\rho}\nonumber\\
&-&\left(f_p\cos\theta+f_s\sin\theta\right)\uvettGreek{\theta}\Big]+\Big[\mathcal{E}_m^{(2)}(\rho)\left(f_p\cos\theta+f_s\sin\theta\right)\nonumber\\
&+&\mathcal{E}_m^{(3)}(\rho)\left(f_s\cos\theta-f_p\sin\theta\right)\Big]\uvettGreek{\zeta}\Big\},
\earr
where 
\barr
\mathcal{E}_m^{(1)}(\rho)&=&D_m\omega(\beta+m\Lambda)\text{J}_m(\alpha_m\rho),\\
\mathcal{E}_m^{(2)}(\rho)&=&D_m(m\omega/\rho)\text{J}_m(\alpha_m\rho),\\
\mathcal{E}_m^{(3)}(\rho)&=&iD_m\omega\left[\alpha_m\text{J}_{m-1}(\alpha_m\rho)-\frac{m}{\rho}\text{J}_m(\alpha_m\rho)\right],
\earr
are the radially dependent expansion coefficients for the electric field
\barr\label{magnetic}
\vett{B}(\vettGreek{\rho}) &=&e^{i(\beta\zeta-\omega t)}\sum_{m\in\mathcal{M}}e^{im\Phi}\Big\{\Big[\mathcal{B}_m^{(1)}(\rho)\left(f_p\cos\theta+f_s\sin\theta\right)\nonumber\\
&+&\mathcal{B}_m^{(2)}(\rho)\left(f_s\cos\theta-f_p\sin\theta\right)\Big]\uvettGreek{\rho}+\Big[\mathcal{B}_m^{(3)}(\rho)\left(f_p\cos\theta+f_s\sin\theta\right)\nonumber\\
&+&\mathcal{B}_m^{(4)}(\rho)\left(f_s\cos\theta-f_p\sin\theta\right)\Big]\uvettGreek{\theta}+\Big[\mathcal{B}_m^{(5)}(\rho)\left(f_p\cos\theta+f_s\sin\theta\right)\nonumber\\
&+&\mathcal{B}_m^{(6)}(\rho)\left(f_s\cos\theta-f_p\sin\theta\right)\Big]\uvettGreek{\zeta}\Big\},
\earr
where
\barr
\mathcal{B}_m^{(1)}(\rho) &=&D_m\left[-\frac{\alpha_m}{\rho}\text{J}_m^{'}(\alpha_m\rho) +2m^2\text{J}_m(\alpha_m\rho)\right],\\
\mathcal{B}_m^{(2)}(\rho) &=&D_m\left[\frac{im\alpha_m}{\rho}\text{J}_m^{'}(\alpha_m\rho)+2im\text{J}_m(\alpha_m\rho)\right],\\  
\mathcal{B}_m^{(3)}(\rho) &=&\frac{imD_m}{\rho}\left[\alpha_m\text{J}_m^{'}(\alpha_m\rho)-\text{J}_m(\alpha_m\rho)\right],\\
\mathcal{B}_m^{(4)}(\rho) &=& D_m\left[(\beta+m\Lambda)^2\text{J}_m(\alpha_m\rho)-\text{J}_m^{''}(\alpha_m\rho)\right],\\
\mathcal{B}_m^{(5)}(\rho) &=&-\frac{mD_m(\beta+m\Lambda)}{2}\text{J}_m(\alpha_m\rho),\\
\mathcal{B}_m^{(6)}(\rho) &=&iD_m\alpha_m(\beta+m\Lambda)\text{J}_m^{'}(\alpha_m\rho),
\earr
are the radially dependent expansion coefficients for the magnetic field.
\section*{Appendix B: Explicit Expression for the Vector Potential for RSABs}
The vector potential can be defined from the Hertz potential as $\vett{A}(\vettGreek{\rho},t)=\nabla\times\vettGreek{\Pi}(\vettGreek{\rho},t)$ \cite{stratton}. Using Eqs. \eqref{eq6} and \eqref{eq6bis} the vector potential for an arbitrary polarised vector RSAB is given, in cylindrical coordinates, as follows:
\barr\label{eqB1}
\vett{A}(\vettGreek{\rho},t) &=& \sum_{m\in\mathcal{M}}D_me^{i[m(\theta+\Lambda\zeta)+\beta\zeta-\omega t]}\Bigg\{-i(\beta+m\Lambda)\text{J}_m(\alpha_m\rho)\Bigg[(f_s\cos\theta-f_p\sin\theta)\,\uvettGreek{\rho}\nonumber\\
&+&i(f_p\cos\theta+f_s\sin\theta)\,\uvettGreek{\theta}\Bigg]\nonumber\\
&+&\Bigg[\alpha_m\left(f_p\cos\theta+f_s\sin\theta\right)\text{J}_m^{'}(\alpha_m\rho)-\frac{im}{\rho}\Big(f_p\cos\theta\nonumber\\
&+&f_s\sin\theta\Big)\text{J}_m(\alpha_m\rho)\Bigg]\,\uvettGreek{\zeta}\Bigg\}.
\earr
For the case of circular polarisation, the above expression simplifies to
\barr\label{eqB2}
\vett{A}(\vettGreek{\rho},t) &=& \sum_{m\in\mathcal{M}}\frac{D_m}{\sqrt{2}}e^{i[m(\theta+\Lambda\zeta)+\sigma\theta+\beta\zeta-\omega t]}\Bigg[A_{\rho}^{(m)}(\rho)\,\uvettGreek{\rho}+A_{\theta}^{(m)}(\rho)\,\uvettGreek{\theta}\nonumber\\
&+&A_{\rho}^{(m)}(\zeta)\,\uvettGreek{\zeta}\Bigg],
\earr
where
\bseq\label{eqB3}
\begin{align}
A_{\rho}^{(m)}(\rho) &=\sigma(\beta+m\Lambda)\text{J}_m(\alpha_m\rho),\\
A_{\theta}^{(m)}(\rho) &=i(\beta+m\Lambda)\text{J}_m(\alpha_m\rho),\\
A_{\zeta}^{(m)}(\rho) &=i\Bigg[\sigma\alpha_m\,\text{J}_m^{'}(\alpha_m\rho)-\frac{m}{\rho}\text{J}_m(\alpha_m\rho)\Bigg].
\end{align}
\eseq

\section*{References}

\end{document}